\documentclass[prb,showpacs,preprintnumbers,twocolumn,amsmath,amssymb]{revtex4}
\usepackage{graphicx}%
\usepackage{color}
\usepackage{bm}
\usepackage{amsmath}
\usepackage{braket}

\begin{document}


\title{Spin-flip transitions and departure from the Rashba model in the Au(111) surface}

\author{Julen Iba\~{n}ez-Azpiroz$^{1,2}$, Aitor Bergara$^{1,2,3}$, E. Ya. Sherman$^{4,5}$, Asier Eiguren$^{1,2}$}
\address{$^{1}$Materia Kondentsatuaren Fisika Saila, Zientzia eta
Teknologia Fakultatea, Euskal Herriko Unibertsitatea, 644
Postakutxatila, 48080 Bilbao, Basque Country, Spain}
\address{$^{2}$Donostia International Physics Center (DIPC), Paseo Manuel de Lardizabal 4, 20018 Donostia/San Sebastian, Spain}
\address{$^3$Centro de F\'{i}sica de Materiales CFM - Materials Physics Center MPC, Centro Mixto CSIC-UPV/EHU,
Edificio Korta,
Avenida de Tolosa 72, 20018 Donostia, Basque Country, Spain}
\address{$^4$Department of Physical Chemistry
University of Basque Country UPV-EHU
48080 Bilbao, Bizkaia, Spain}
\address{$^5$IKERBASQUE, Basque Foundation for Science,
Bilbao, Spain}

\date{\today}

\pacs{71.70.-d, 72.25.Rb, 73.21.-b
}

\begin{abstract}
We present a detailed analysis of the spin-flip 
excitations induced by a periodic time-dependent electric field 
in the Rashba prototype Au(111) noble metal surface.
Our calculations incorporate the full spinor structure of the spin-split 
surface states and employ a Wannier-based scheme for the spin-flip matrix elements.
We find that the spin-flip excitations associated with the surface states exhibit  
an strong dependence on the electron momentum magnitude, a feature that is
absent in the standard Rashba model~\cite{rashba}.
Furthermore, we demonstrate that the maximum of the calculated 
spin-flip absorption rate is  
about twice the model prediction.
These results show that although the Rashba model accurately describes
the spectrum and spin polarization, it does not fully account for the
dynamical properties of the surface states.

\end{abstract}
\maketitle

\section{INTRODUCTION}
\label{sec:intro}

Surfaces represent an ideal testing 
ground for investigating the nature 
of the relativistic spin-orbit interaction in low dimensional 
systems~\cite{so-bihlmayer,spin-interf-1,spin-interf-2}.
As pointed in the pioneering work 
by LaShell \textit{et al.}~\cite{lashell}, 
the lack of inversion symmetry of surfaces
allows for the spin-splitting of the 
Shockley-type surface states via the spin-orbit interaction.
Noteworthy, the order of magnitude of the energy spin-splitting
is as much as $\sim$0.1-0.5 eV, about two orders of magnitude larger than 
in semiconductors~\cite{so-bihlmayer}.
Recently, several studies performed on coated surfaces have
revealed exceptional effects of the spin-orbit interaction. 
The family of bismuth alloy surfaces~\cite{bi-te-etxenike,ast_giant_2007,bi100}, for instance, 
exhibits giant spin-orbit energy 
shifts of nearly $400$ meV.
Other interesting examples 
include the semiconducting surfaces
Tl/Si(111)$-(1\times1)$~\cite{abrupt,minghao,tlsi111}, 
Tl/Ge(111)$-(1\times1)$~\cite{tl-ge111} and 
\cite{scontr,pbge111-spin-flip} Pb/Ge(111)$-\beta\sqrt{3}\times\sqrt{3}R30\,^{\circ}$,
among many others. In these systems, the bulk bands present a gap near the Fermi level, 
and thus, the electron transport properties are strongly influenced by the spin-split 
metallic surface states. Even surfaces with light element over-layers 
such as H/W(110)$-(1\times1)$ reveal extremely complex spin polarization structure,  
which is inherent of the anisotropy of the spin-orbit interaction~\cite{hw111-exp,hw111-calc}.

A particularly appealing aspect about surfaces 
is the possibility of manipulating the electron spin 
by means of externally applied electric 
fields~\cite{datta,PhysRevLett.91.126405,rashba-sheka,PhysRevB.74.165319,PhysRevB.85.125312,PhysRevB.70.201309}.
The basic idea in this scenario
would be to control
the spin orientation by inducing spin-flip 
excitations between the spin-split surface states.
In practice, this is done by applying an external electric field which 
couples to the spin-dependent electron velocity due to the spin-orbit interaction.
Since electric fields are easily created and manipulated experimentally, the mentioned mechanism 
(electric dipole spin resonance)
could offer new perspectives for future applications in spintronics.

In this paper, we present fully relativistic 
first-principles calculations for analyzing
the spin-flip excitations induced by time-dependent electric fields 
in the Au(111) noble metal surface. 
This system is considered as the paradigm of a two-dimensional 
free electron-like gas with the  
Rashba-type spin-orbit coupling~\cite{lashell,rashba}. 
It is commonly believed that the properties of the Au(111) surface states,
such as the energy spin-splitting or the spin polarization structure are well
described in terms of the Rashba model~\cite{so-bihlmayer,au111-exp,PhysRevB.78.195413,PhysRevB.68.165416,premper}.
In fact, the noble metal (111) surfaces have been considered as an almost perfect realization 
of the Rashba Hamiltonian.
However, we demonstrate in this work that the spin related response properties of the Au(111) surface states
show a detectable departure from this model. 
In particular, our calculations demonstrate that the spin-flip transition  
probability reveals an appreciable angular and momentum dependence, while in the Rashba model this
quantity appears with a trivial functional shape.
Furthermore, we find that the maximum value of the calculated spin-flip absorption rate is  
almost double  the model prediction.

\section{THEORETICAL FRAMEWORK}
\label{sec:theory}

In this section, we briefly introduce the computational approach   
for analyzing the spin-flip excitations 
associated with the spin-split surface states.
Unless otherwise stated, atomic units will be used throughout the work, 
$e=\hbar=m_e=4\pi\epsilon_0=1$.
Let us start by considering the following single-particle Hamiltonian
including the spin-orbit interaction,
\begin{equation}\label{eq:h0}
\hat{\text{H}}_{0}=\frac{\hat{\textbf{p}}^{2}}{2}-V(\hat{\textbf{r}})-\frac{1}{4c^{2}}\hat{\boldsymbol{\sigma}}\cdot\left(\hat{\boldsymbol{\nabla}}V(\hat{\textbf{r}})\times{\hat{\textbf{p}}}\right),
\end{equation} 
where $V(\hat{\textbf{r}})$ is the scalar potential and 
$\hat{\boldsymbol{\sigma}}$ represent the Pauli spin operator.
In this framework, we analyze the response of an electron state to an 
external time-dependent electric field described by the vector potential
$\hat{\textbf{A}}_{\text{ext}}(t)=\textbf{A}^{(\alpha)}_{0}
\cos\omega t$,
where $\omega$ and $\alpha$ represent the frequency
and polarization of the external field respectively. 
The spatial variation of the field is neglected since we are interested
in the optical limit ($\textbf{q}\rightarrow0$), in which case
the field can be considered as spatially constant~\cite{pbge111-spin-flip,sherman_minimum}.

We adopt the following conventions for the 
$x$ and $y$ linearly polarized, $\textbf{A}^{(x)}_{0}=A_{0}\hat{\textbf{x}}$ and
$\textbf{A}^{(y)}_{0}=A_{0}\hat{\textbf{y}}$,  and the 
right (\textit{R}) and left (\textit{L}) circularly polarized light,
$\textbf{A}_{0}^{(R,L)}=A_{0}(\hat{\textbf{x}}\pm i\hat{\textbf{y}})/\sqrt{2}$
(see Fig. \ref{fig:electron-structure}a for the axes convention).

In our perturbation theory treatment, the leading term describing the 
interaction between the external field and the electron gas appears as ~\cite{PhysRevLett.91.126405} 
\begin{eqnarray}\label{eq:hint}
\hat{\text{H}}_{\text{int}}(t)&=& 
- \frac{1}{c}\hat{\textbf{v}}\cdot\hat{\textbf{A}}_{\text{ext}}(t),
\end{eqnarray}
where $\hat{\textbf{v}}$ represents the electron velocity operator, commonly expressed as
\begin{eqnarray}\label{eq:vc}
\hat{\textbf{v}}= -i[\hat{\textbf{r}},\hat{\text{H}}_{\text{0}}]=
\dfrac{\partial\hat{\text{H}}_{\text{0}}}{\partial \hat{\textbf{p}}}=
\hat{\textbf{p}} - 
\dfrac{1}{4c^{2}}(\hat{\boldsymbol{\sigma}}\times\hat{\boldsymbol{\nabla}}V(\hat{\textbf{r}})).
\end{eqnarray}
We observe that apart from the canonical contribution $\hat{\textbf{p}}$, 
the velocity operator contains an additional term which directly depends on the spin.
It is precisely due to this term in Eq. \ref{eq:vc} that the interaction term in Eq. \ref{eq:hint}
is allowed to produce spin-flip transitions 
among spin-split surface states.

The transition rate associated to $\hat{\text{H}}_{\text{int}}(t)$
is calculated considering the ordinary 
first order perturbation (Fermi's golden rule), 
%
\begin{equation}\label{eq:tr-rate}
\begin{split}
\gamma^{(\alpha)}_{m  n  }(\omega)=&2\pi\int 
\left( f_{{\bf k}m}-f_{{\bf k}n} \right)
 |C^{(\alpha)}_{m n}(\textbf{k})|^{2} \\
&\times \delta(\epsilon_{{\bf k}n}-\epsilon_{\textbf{k}m}-\omega)\frac{d^{2}k}{(2\pi)^{2}}
.
\end{split}
\end{equation}
The above describes transitions from state $m$ to $n$, with
$f_{{\bf k}i}$ and $\epsilon_{{\bf k}i}$ 
the Fermi-Dirac distribution function and surface state eigenvalue, respectively, with $i=m,n$.

The velocity operator, which is the origin of the electron spin-flip transitions, 
enters the matrix elements in Eq. \ref{eq:tr-rate} as
\begin{equation}\label{eq:sf-mat-elem-1}
\begin{split}
C^{(\alpha)}_{m n  }(\textbf{k})=-\dfrac{1}{2c}\textbf{A}^{(\alpha)}_{0}\cdot\bra{\Psi_{\textbf{k} m}} \hat{\textbf{v}}
\ket{\Psi_{{\bf k}n}},
\end{split}
\end{equation}
where $\Psi_{\textbf{k} i}(\textbf{r})=e^{i\textbf{k}\cdot\textbf{r}}\cdot u_{\textbf{k} i}(\textbf{r})$ represents  
the single-particle Bloch spinor wave function of surface states. 

Considering the Ehrenfest theorem for the velocity operator, 
$\hat{\textbf{v}}=-i[\hat{\textbf{r}},\hat{\text{H}}_{\text{0}}]$, and 
expanding the commutator $[\hat{\textbf{r}},\hat{\text{H}}_{\text{0}}]$, 
the matrix elements of Eq. \ref{eq:sf-mat-elem-1} can be cast 
into the following form~\cite{blount,k-gradients},
\begin{equation}\label{eq:sf-mat-elem-2}
\begin{split}
C^{(\alpha)}_{m n  }(\textbf{k})=-\dfrac{\epsilon_{{\bf k}n}-\epsilon_{\textbf{k} m}}{2c}\textbf{A}^{(\alpha)}_{0}\cdot 
\bra{u_{\textbf{k} m}} \hat{\boldsymbol{\nabla}}_{\textbf{k}}\ket{u_{{\bf k}n}}
.
\end{split}
\end{equation}
The $\bra{u_{\textbf{k} m}} \hat{\boldsymbol{\nabla}}_{\textbf{k}}\ket{u_{{\bf k}n}}$ matrix element
is precisely the generalized (non-Abelian) Berry connection~\cite{berry}
associated to the spin-split surface states.

It is noteworthy that the spin non-collinearity is the ultimate reason why the above term does not vanish, as
briefly illustrated in the next lines.
Let us begin by considering a system without spin-orbit coupling 
and subjected to a constant magnetic field along the $z$ axis. 
In these conditions, the spinor states would be collinearly 
polarized along the $z$ axis, i.e., we would have 
$u_{\textbf{k} n}(\textbf{r})=g_{\textbf{k} n}(\textbf{r})
\begin{footnotesize}
\left(\begin{matrix}
  1\;,\; 0
\end{matrix}\right)^{\text{T}}
\end{footnotesize}
$ 
and 
$u_{\textbf{k} m}(\textbf{r})=g_{\textbf{k} m}(\textbf{r})
\begin{footnotesize}
\left(\begin{matrix}
  0\;,\;
  1
\end{matrix}\right)^{\text{T}}
\end{footnotesize}
$, where $\text{T}$ stands for matrix transposition. 
Since the momentum operator 
$\hat{\boldsymbol{\nabla}}_{\textbf{k}}$ is diagonal in the spin-basis, 
one deduces that the matrix elements entering 
Eq. \ref{eq:sf-mat-elem-2} vanish identically. 
This is in complete 
contrast to the situation when a finite spin-orbit interaction is 
present. This interaction induces
an explicit momentum dependence of the spinor wave function,
$u_{\textbf{k} i}(\textbf{r})=
\begin{footnotesize}
\left(\begin{matrix}
  g^{+}_{\textbf{k} i}(\textbf{r})\;,\;
  g^{-}_{\textbf{k} i}(\textbf{r})
\end{matrix}\right)^{\text{T}}
\end{footnotesize}$.
In this case, it is obvious that the spinor $\hat{\boldsymbol{\nabla}}_{\textbf{k}}u_{\textbf{k} n}(\textbf{r})=
\begin{footnotesize}
\left(\begin{matrix}
  \hat{\boldsymbol{\nabla}}_{\textbf{k}}g^{+}_{\textbf{k} n}(\textbf{r})\;,\;
  \hat{\boldsymbol{\nabla}}_{\textbf{k}}g^{-}_{\textbf{k} n}(\textbf{r})
\end{matrix}\right)^{\text{T}}
\end{footnotesize}
$
does not generally describe a spin orientation parallel to the original spinor $u_{\textbf{k} n}(\textbf{r})$.
Thus, an appreciable magnitude of the 
$\bra{u_{\textbf{k} m}} \hat{\boldsymbol{\nabla}}_{\textbf{k}}\ket{u_{{\bf k}n}}$
matrix elements entering Eq. \ref{eq:sf-mat-elem-2} is a direct 
consequence of the spin noncollinearity introduced by the spin-orbit interaction. 
The opposite is also true and 
one deduces that 
if the direction of the spin polarization experiences a 
significant variation in some \textbf{k}-space region, 
the associated spin-flip matrix elements would 
accordingly be enhanced,
as found in a previous analysis~\cite{pbge111-spin-flip}. 

The calculation of the momentum gradient in Eq. \ref{eq:sf-mat-elem-2}
presents a computational challenge because of the 
inherent phase indeterminacy carried by
the spinor wave-functions~\cite{blount}. 
As a consequence, simple finite difference formulas cannot be directly applied, 
and a gauge fixing procedure is needed.
Recently, a new method 
to solve this problem has been presented ~\cite{k-gradients,wannier-2012-lopez} whereby 
the matrix elements are re-expressed in terms of the 
maximally localized Wannier functions~\cite{wannier90}.
In this scheme, the maximal localization of the Wannier states amounts to
obtain the smoothest possible variation of the spinor states within the Brillouin zone,
hence the interpolation procedure is optimized. 
Following this approach, the $C^{\alpha}_{m n}(\textbf{k})$ matrix elements entering Eq. \ref{eq:tr-rate} 
can be interpolated into a very fine \textbf{k} mesh
with a negligible computational cost~\cite{k-gradients,fermi-surf-wann-wang,wannier-2012-lopez}.
In the present work, the spin-flip matrix elements and surface state eigenvalues
entering the integral of Eq. \ref{eq:tr-rate}
have been evaluated in a dense $1000\times1000$ \textbf{k}-point grid.
This has allowed to consider a very fine Gaussian width of $4\cdot10^{-4}$ eV 
for the integral. 

\begin{figure}[t]
\includegraphics[width=0.35\textwidth]{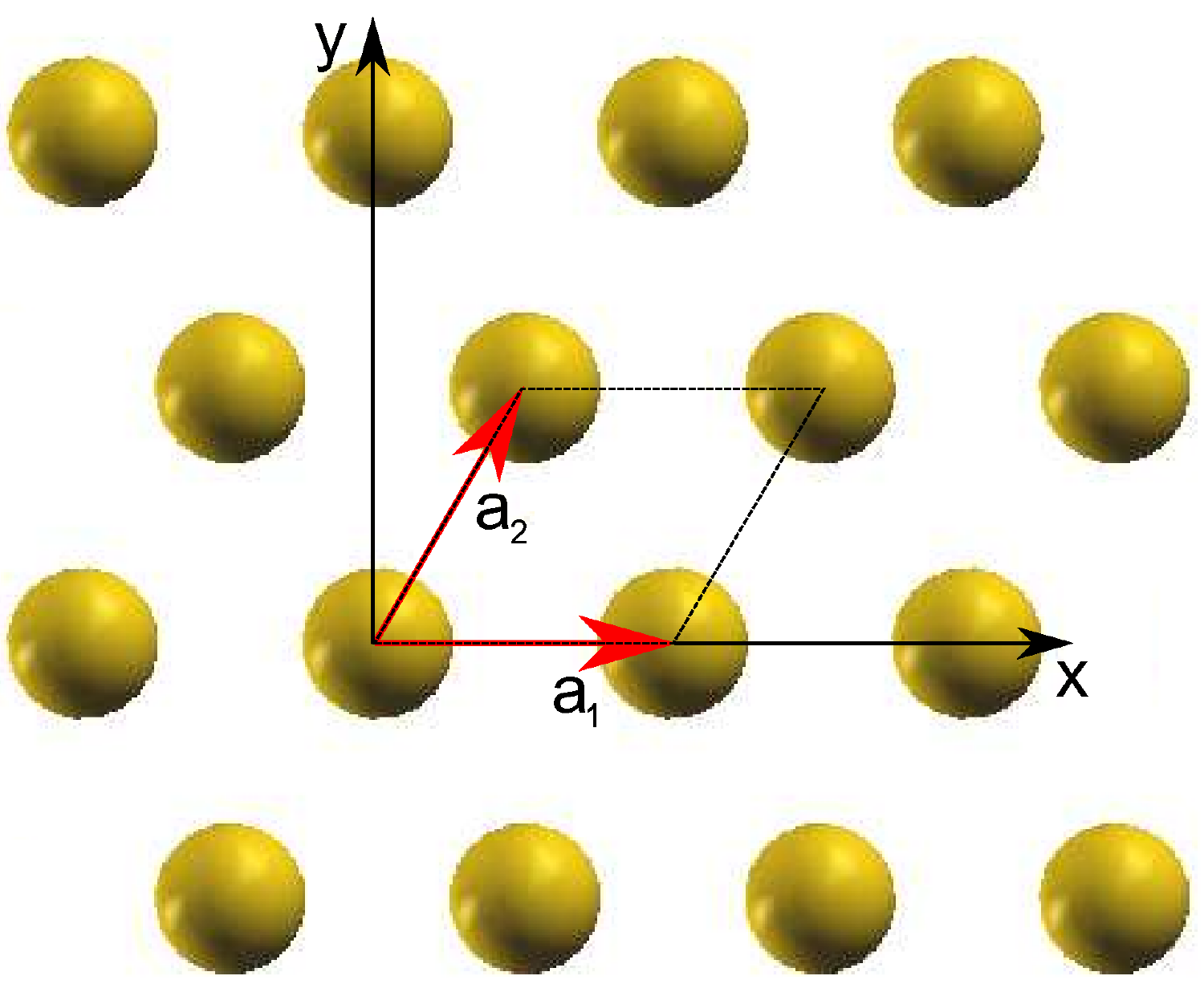}
\caption{(color online) Top view of the Au(111) surface.~\cite{xcrysden} 
Gold atoms are represented by the
spheres (yellow). The big (red) arrows denote the
2D direct lattice defined in the text.}
\label{fig:au-structure}
\end{figure}

The ground state electronic properties of the Au(111) surface
have been calculated within the non-collinear LDA-DFT 
formalism, considering a plane wave basis as implemented in the 
QUANTUM ESPRESSO package~\cite{espresso}.
We used a plane wave cutoff corresponding to $E_c=$55 Ry and a $32\times32$ Monkhorst-Pack mesh~\cite{MParck}
for the self-consistent cycle.
The spin-orbit interaction has
been incorporated considering  
norm-conserving fully relativistic pseudo-potentials for the 5d and 6s electrons of Au,
including the full spinor structure of the wave functions~\cite{Dalcorso}.
The Au(111) surface was modeled by the repeated
slab technique considering 21 Au layers,
with 21 {\AA{ 
of vacuum separating the two sides of the slab
and $\sim2.49$ {\AA} 
separating adjacent Au layers (interlayer spacing).
In this surface, a single 2D lattice is described by the basis vector 
$\textbf{a}_{1}=a\hat{\textbf{x}}$,  
$\textbf{a}_{2}=(\hat{\textbf{x}}+\sqrt{3}\hat{\textbf{y}})a/2$,
with associated reciprocal basis 
vectors $\textbf{b}_{1}=2\pi/a(\hat{\textbf{x}}-\hat{\textbf{j}}/\sqrt{3})$,
$\textbf{b}_{2}=2\pi/a\cdot2\hat{\textbf{y}}/\sqrt{3}$,
where $a=3.12$ \AA.
We have included the top view of the surface in Fig. \ref{fig:au-structure}.
In our optimized configuration, all forces acting on individual atoms were smaller than $1.8 \cdot10^{-4}$ Ry$\cdot$\AA$^{-1}$.
\begin{figure}[t]
\includegraphics[width=0.47\textwidth]{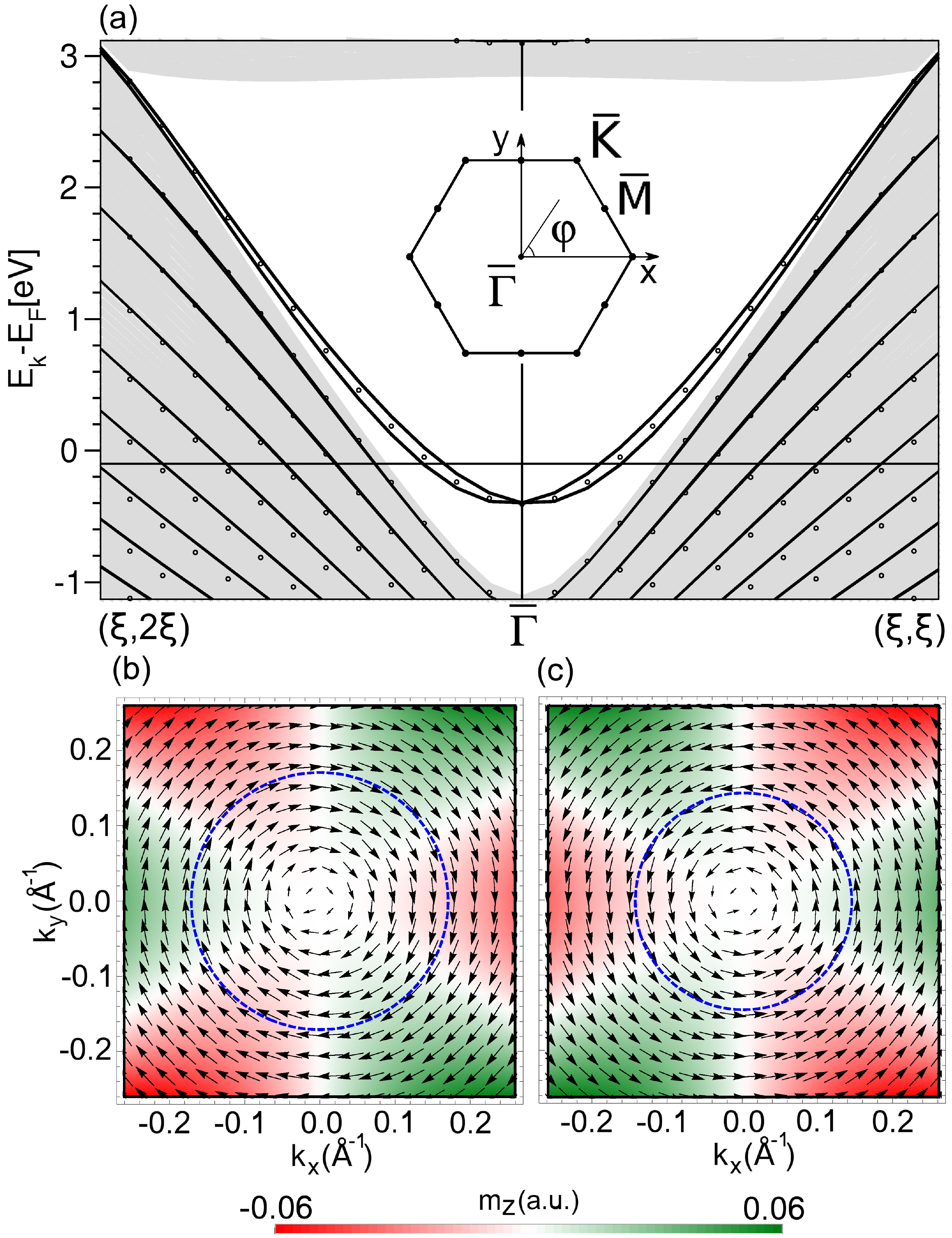}
\caption{(color online) 
(a) Band structure of the Au(111) surface close to the $\overline{\Gamma}$ point
($\xi=0.25$ $2\pi/a$). 
The scalar relativistic and fully relativistic bands are represented by circles 
and solid lines, respectively. 
The continuous background (grey) denotes the bulk band projection.
(b) and (c) Ab-initio momentum dependent spin polarization 
associated to lower and higher spin-split sub-bands, respectively. 
Arrows represent the in-plane spin-polarization component, 
whereas the background color code indicates the surface perpendicular component, $m_{i,z}(\textbf{k})$.
Dashed (blue) lines indicate the calculated ab-initio Fermi surface associated 
with each surface sub-band. The radii of the circles in (b) and (c) are
$k_{F}^{-}$ and $k_{F}^{+}$, respectively.
}
\label{fig:electron-structure}
\end{figure}

\section{RESULTS}
\label{sec:results}

\subsection{Ground state properties}
\label{sec:ground-state}

In Fig. \ref{fig:electron-structure}a we present the calculated
electron band structure of the the Au(111) surface.
The scalar relativistic (without spin-orbit interaction)
and fully relativistic bands correspond to
circles and solid lines, respectively. The bulk
band projection is indicated by the background continuum (grey). 
While the scalar relativistic calculation shows a single spin-degenerate surface band
outside the bulk band projection,
the fully relativistic calculation shows
the two well known spin-split metallic surface state bands
measured for the first time by LaShell \textit{et al.}~\cite{lashell}
We observe that far from $\overline{\Gamma}$, the spin-split surface 
bands gradually spin-degenerate
as they approach the bulk
projection (continuum) and become resonance states.
The calculated binding energy  
at $\overline{\Gamma}$ is 420 meV, 
while the spin-splitting 
at the Fermi level ranges approximately from 120 to 135 meV,
corresponding to the Fermi wave vectors $k_{F}^{+}=0.145$ $\text{\AA}^{-1}$
and $k_{F}^{-}=0.175$ $\text{\AA}^{-1}$, respectively. 

\begin{figure}[t]
\centering
\includegraphics[width=0.4\textwidth]{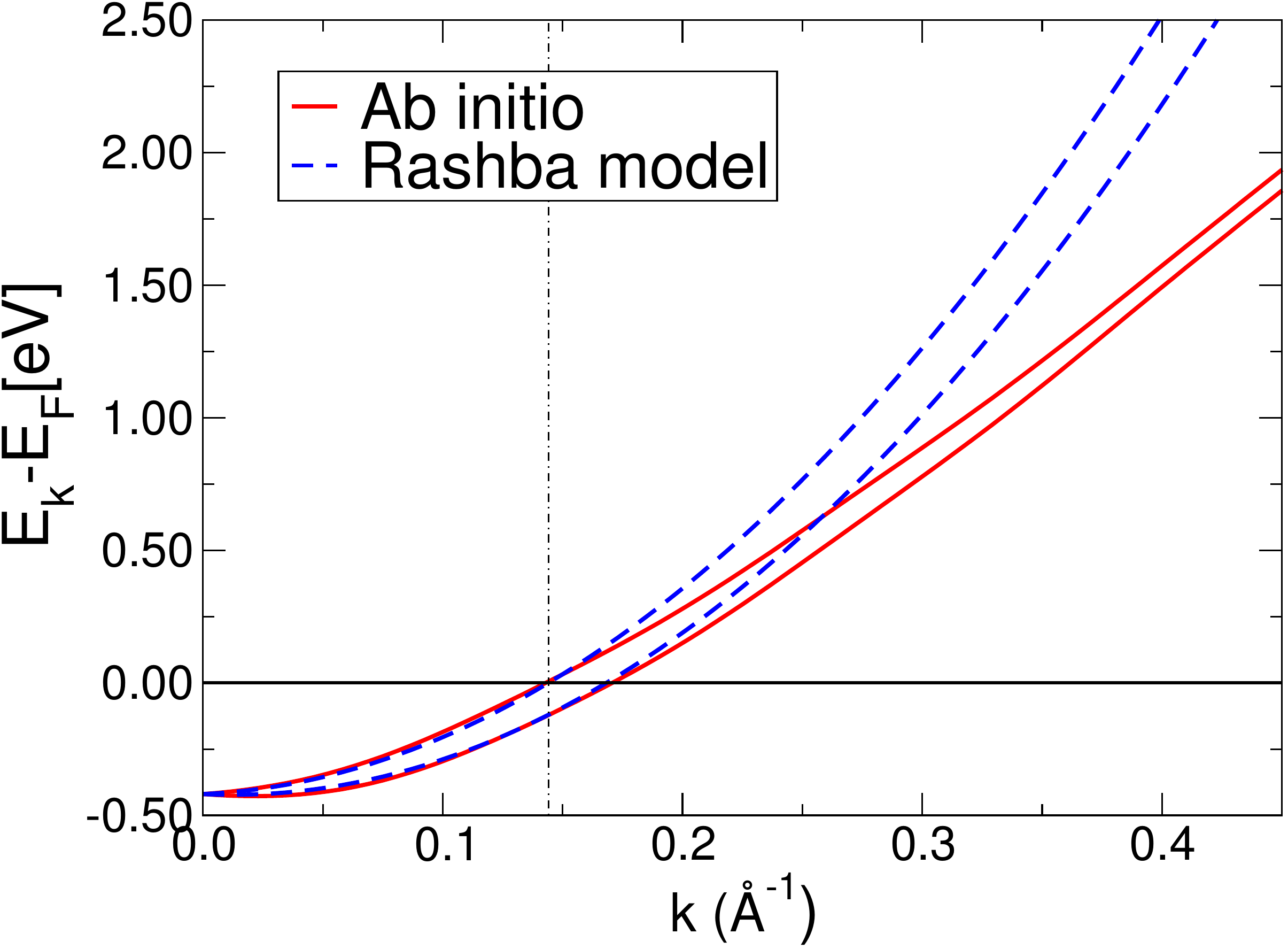}
\caption{(color online) Comparison between the Rashba model prediction for the energy 
dispersion (Eq. \ref{eq:rashba-energy}) and the calculated ab-initio 
band structure of the surface states.
The Rashba parameter $\alpha_{R}$ has been fitted at $k_{F}^{+}=0.145$ $\text{\AA}^{-1}$, indicated by the
vertical line.
}
\label{fig:au111-rashba-bands}
\end{figure}

It is instructive to compare the calculated band structure of the surface states
with the Rashba model energy dispersion of Eq. \ref{eq:rashba-energy}.
This model equation predicts a spin-splitting that grows linearly with 
the magnitude of the electron momentum, 
$\Delta E=2|\textbf{k}|\alpha_{R}$, with $\alpha_{R}$ the Rashba parameter.
An explicit value for this parameter can be obtained by extracting
$\Delta E$ and $|\textbf{k}|$ from the ab-initio band structure. 
Since we are interested in the details close to the Fermi level,
we consider the calculated Fermi wave vector, $|\textbf{k}|=k_{F}^{+}=0.145$ $\text{\AA}^{-1}$, 
and the corresponding energy
spin-splitting, $\Delta E=0.12$ eV, obtaining $\alpha_{R}=\Delta E/2|\textbf{k}|=0.419$ eV$\cdot\text{\AA}$.
This value is in good agreement with the one reported in 
a recent ARPES experiment~\cite{au111-exp}, 
$\alpha_{R}=0.396$ eV$\cdot\text{\AA}$. 
Besides, we obtain an effective mass of $m^{*}=0.23$ 
(see Eq. \ref{eq:rashba-energy}) from a parabolic fit to the 
band structure, which also agrees well with experiments~\cite{au111-exp}, $m^{*}=0.25$.

In Fig. \ref{fig:au111-rashba-bands} we compare the
ab-initio band structure of the surface states with the
Rashba model energy dispersion (Eq. \ref{eq:rashba-energy})
calculated using the parameter values 
$\alpha_{R}=0.419$ eV$\cdot\text{\AA}$ and $m^{*}=0.23$.
This figure shows a good agreement for energies below the Fermi level
and, as expected, 
the ab-initio and Rashba model energies  coincide exactly at $|\textbf{k}|=k_{F}^{+}$.
For $|\textbf{k}|>k_{F}^{+}$, the ab-initio energy spin-splitting of the surface states 
ceases to grow linearly and, furthermore, it starts decreasing.
Therefore, the Rashba model energy dispersion clearly deviates from the ab-initio
band structure for $|\textbf{k}|>k_{F}^{+}$.

Figs. \ref{fig:electron-structure}b and \ref{fig:electron-structure}c 
illustrate the momentum dependent spin polarization 
of the spin-split surface states,
\begin{equation}\label{eq:spinpol} 
\boldsymbol{m}_{i}({\bf k})=\int \Psi^{*}_{{\bf k}i}({\bf r})  
\hat{\boldsymbol{\sigma}} \Psi_{{\bf k}i}({\bf r})d^{3}r.
\end{equation}
In the figures, arrows represent the in-plane spin-polarization component, 
while the background code indicates the surface perpendicular component, $m_{z,i}(\textbf{k})$.

Both surface states are spin-polarized 
in practically the opposite direction in agreement with spin-resolved 
ARPES measurements~\cite{lashell,au111-exp},
and describe a circular spin structure around 
the $\overline{\Gamma}$ point following the Rashba
model (Eq. \ref{eq:rashba-spinpol}).
Our calculations confirm that 
$\boldsymbol{m}_{i}(\textbf{k})$ is almost parallel
to the surface for $|\textbf{k}|\lesssim k_{F}^{+}$,
which is the region where the Rashba model is expected to 
properly describe the properties of the surface states.
Instead, the calculated surface-perpendicular 
component ($m_{z,i}(\textbf{k})$) 
acquires a finite value for  $|\textbf{k}|\gtrsim k_{F}^{+}$, 
indicating a departure from the Rashba model in this region.
As shown by Henk \textit{et al.}~\cite{PhysRevB.68.165416},
this feature is a consequence of
in-plane components of the potential gradient associated with the real surface structure.
In our calculations, we find that at $k_{F}^{+}$, the surface-perpendicular
component represents the $\sim3\%$ of the total magnitude of the spin polarization.

\subsection{Spin-flip transitions}
\label{sec:spin-flip}

\begin{figure*}[t]
\includegraphics[width=0.85\textwidth]{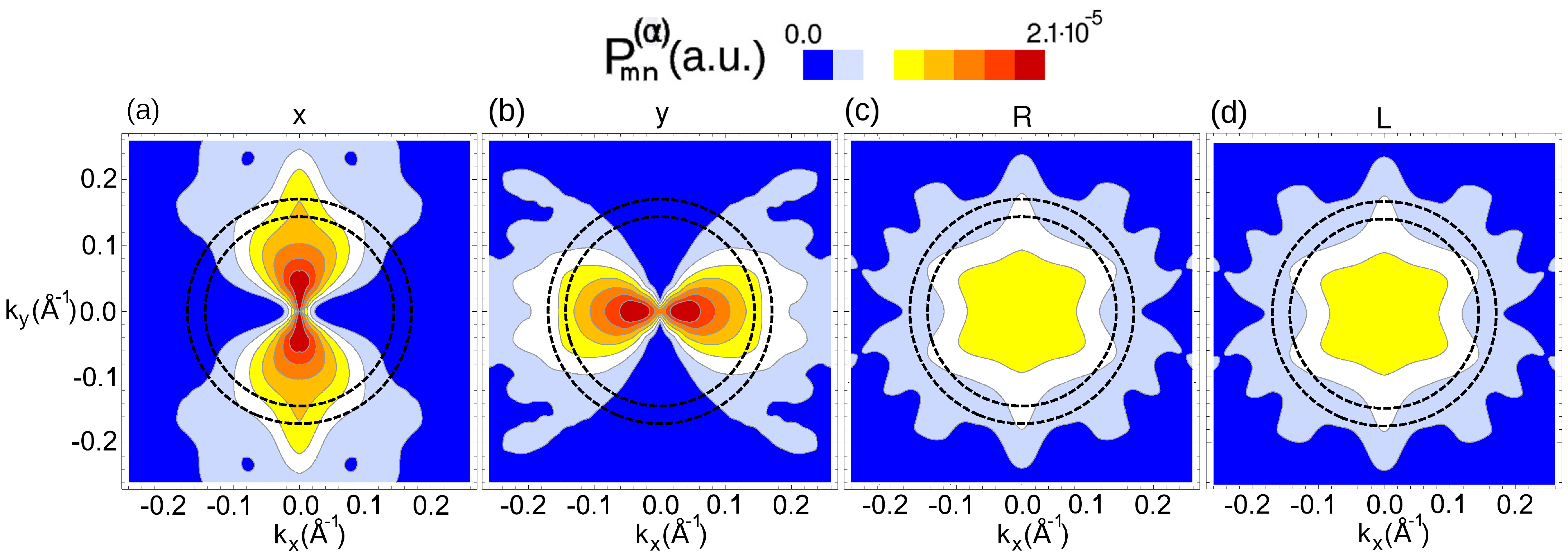}
\caption{(color online)  
(a), (b), (c) and (d) 
Contour plot of the calculated momentum dependent surface state spin-flip transition probability,
associated to the incoming light $\alpha=x, y, R$ and $L$ polarizations, respectively.
The dashed lines (black) 
denote the ab-initio calculated Fermi surface 
associated with the surface states.
The radii of the inner and outer circles are $k_{F}^{+}$ and $k_{F}^{-}$, respectively.
}
\label{fig:au-spin-flip}
\end{figure*}

Figs. \ref{fig:au-spin-flip}a, \ref{fig:au-spin-flip}b, \ref{fig:au-spin-flip}c and \ref{fig:au-spin-flip}d present the 
calculated momentum dependent spin-flip transition probability 
associated with the surface states,
\begin{equation}\label{eq:transition-prob}
P^{(\alpha)}_{mn}(\textbf{k})\equiv \dfrac{|C_{m n  }^{(\alpha)}(\textbf{k})|^{2}}{|\textbf{A}^{(\alpha)}_{0}|^{2}}
,
\end{equation} 
for linearly ($\alpha=x,y$) as well as for 
circularly ($\alpha=R,L$) polarized light, respectively (see Eq. \ref{eq:sf-mat-elem-1}).

In contrast to the Rashba model predicting a constant and totally isotropic
transition probability for circularly polarized 
light (Eq. \ref{eq:rashba-RL-mat-elem}), 
our calculations presented in 
Figs. \ref{fig:au-spin-flip}c and \ref{fig:au-spin-flip}d
describe an appreciable hexagonal angular dependence
inherited from the \textit{C}$_{3}$ symmetry of the real surface structure.
For \textit{x} and \textit{y} linearly polarized light, 
our calculated spin-flip transition probability is in qualitative 
agreement with the dipole-like function predicted by the Rashba model
(Eqs. \ref{eq:rashba-x-mat-elem} and \ref{eq:rashba-y-mat-elem}),
but showing again an appreciable modulation.

Noteworthy, our ab-initio calculations show a clear 
deviation from the Rashba model in one more important aspect;
the dependence of the calculated spin-flip transition probability 
on the momentum magnitude $|\textbf{k}|$.
This feature is particularly evident for the
\textit{x} and \textit{y} linearly polarized light 
(Figs. \ref{fig:au-spin-flip}a and \ref{fig:au-spin-flip}b), 
but it is also present 
for the \textit{R} and \textit{L} circular polarizations
(Figs. \ref{fig:au-spin-flip}c and \ref{fig:au-spin-flip}d). 
In all these cases, the
spin-flip transition probability diminishes with
increasing momentum, a feature that is absent in the ideal Rashba model.
This can be understood as the surface bands
approaching the bulk continuum lose gradually their surface character.

\begin{figure}[b]
\includegraphics[width=0.4\textwidth]{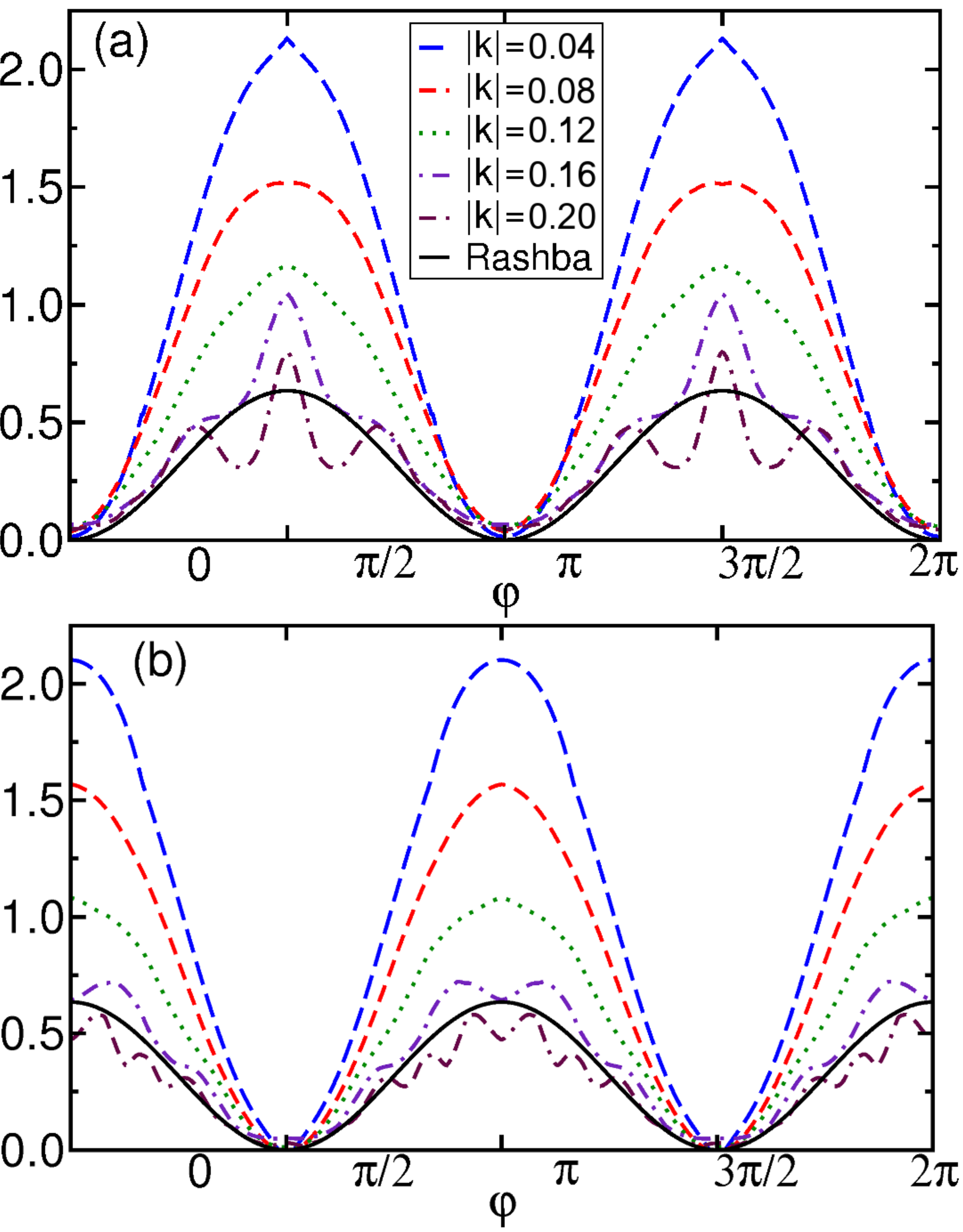}
\caption{(color online) 
Calculated angular dependence of the spin-flip transition probability 
(units of $10^{-5}$ a.u.)
for 
several circular paths corresponding to different absolute values of the electron momentum, $|\textbf{k}|$
(units of \AA$^{-1}$).
Panels (a) and (b) show the results for the incoming $x$ and $y$ linearly polarized light respectively.
The solid (black) lines are the $|\textbf{k}|$-independent Rashba model predictions
of Eqs. \ref{eq:rashba-x-mat-elem} and \ref{eq:rashba-y-mat-elem}, with 
the parameter $\alpha_{R}$ extracted from the 
ab-initio band structure at $k_{F}^{+}$.
}
\label{fig:au-sflip-circular-x-y}
\end{figure}

In Figs. \ref{fig:au-sflip-circular-x-y}a and 
\ref{fig:au-sflip-circular-x-y}b we explicitly analyze the angular dependence of the probability distribution
for the $x$ and $y$ linearly polarized light, 
following several circular paths centered at high symmetry point $\overline{\Gamma}$ with fixed momentum $|\textbf{k}|$. 
We observe that the calculated $P^{(x)}_{mn}(\textbf{k})$ and $P^{(y)}_{mn}(\textbf{k})$
closely follow the dipole-like functional shape of the Rashba model
$(\sin^{2}\varphi, \cos^{2}\varphi)$,
specially for small momenta, $|\textbf{k}|\lesssim k_{F}^{+}$. 
We find that even though the order of magnitude coincides for all $|\textbf{k}|$,
the calculated spin-flip transition probability 
shows a remarkable modulation with respect to the Rashba model result
near $k_{F}^{+}$.

\begin{figure}[b]
\includegraphics[width=0.4\textwidth]{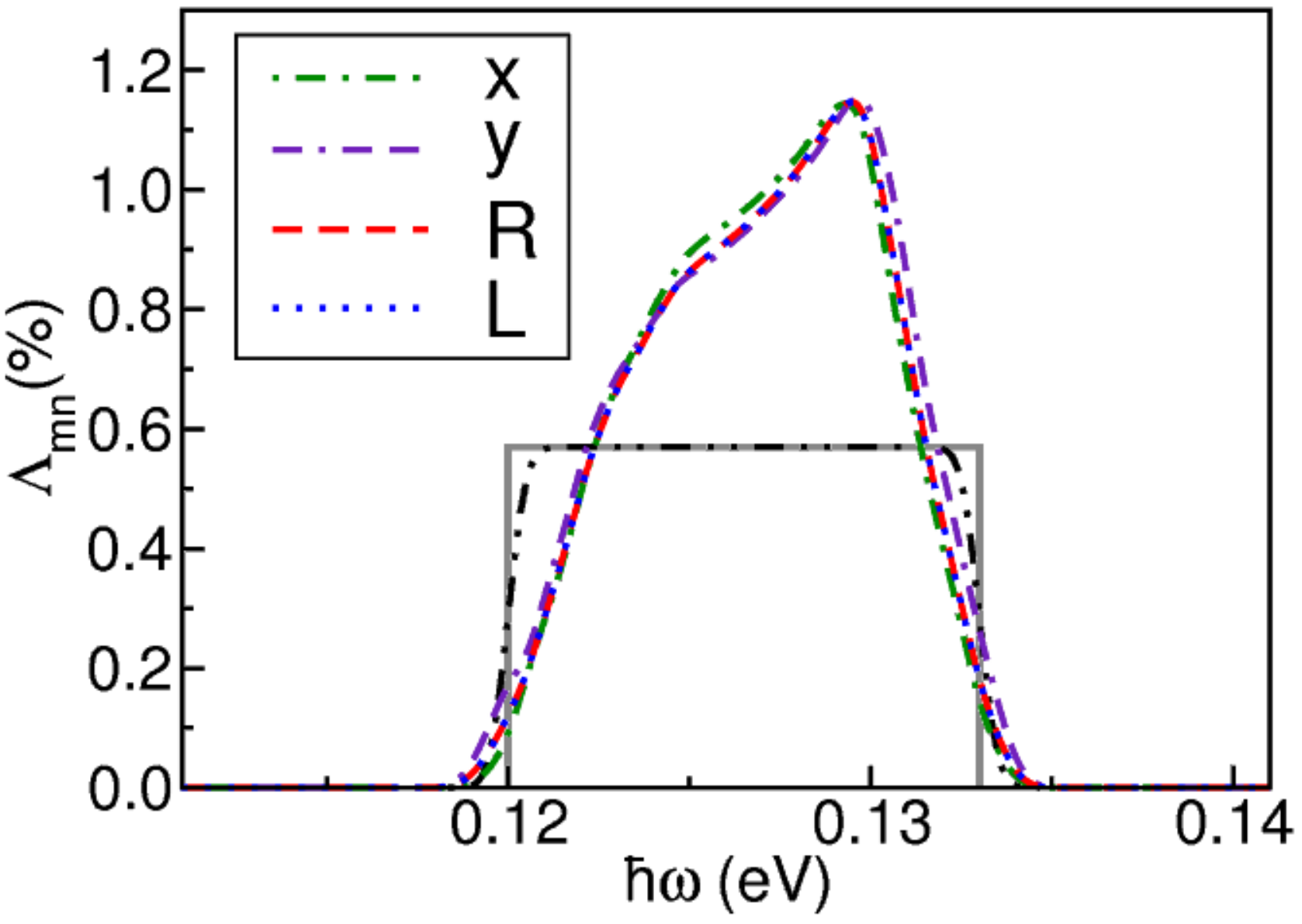}
\caption{(color online) Integrated spin-flip absorption rate for the Au(111) surface.
Super-imposed dashed (red), dotted (blue), dot-dashed (green) and dot-dash-dashed (purple) lines 
represent the calculated results corresponding to the
$x$ and $y$ linear and $R$ and $L$ circularly polarized light, respectively.
The constant solid (grey) and dot-dot-dashed (black) lines denote respectively the Rashba model prediction without and
with broadening.
}
\label{fig:au-abs}
\end{figure}

In Fig. \ref{fig:au-abs} we present the calculated 
absorption rate associated to the spin-flip excitations,
\begin{eqnarray}
\Lambda^{(\alpha)}_{mn  }(\omega)=
\dfrac{\omega\cdot\gamma^{(\alpha)}_{mn  }(\omega)}{\mathcal P},
\end{eqnarray}
where $\gamma^{(\alpha)}_{m n}(\omega)$ is the spin-flip transition rate
of Eq. \ref{eq:tr-rate}, and ${\mathcal P}=|\textbf{A}^{(\alpha)}_{0}|^{2}\omega^{2}/8\pi c$ is
the optical power per unit area of the incident field.
Thus, $\Lambda^{(\alpha)}_{mn  }(\omega)$ measures the percentage of the total irradiated light absorbed 
in the spin-flip processes.

We do not find any significant difference among the $x$ and $y$ linear polarization because
of the isotropy of the problem, as the orbital components of the surface states in Au(111) are mainly 
$s$, $p_{z}$ and $d_{z^{2}}$~\cite{au111-exp,PhysRevB.78.195413}.
The results for the $R$ and $L$ circular polarizations
superimpose due to the presence of the time reversal symmetry ~\cite{dichroism-souza,valleytronics-prb}.
The solid (grey) line indicates the constant Rashba model prediction
(for zero temperature),
$\Lambda_{-+}=\pi/4c
\sim 0.57\%$, 
which is independent of the external field 
polarization and even of the Rashba parameter $\alpha_{R}$.
We deduce from Fig. \ref{fig:au-abs} that  
light is absorbed in the 120-135 meV energy window, 
corresponding to the range of the calculated energy 
spin-splitting of the surface bands close to the Fermi level.
Fig. \ref{fig:au-abs} reveals also that
the Rashba model underestimates the maximum magnitude 
of the spin-flip absorption
rate by approximately 100\%. 
The reason is that within the Brillouin zone area 
where spin-flip excitations are allowed 
($k_{F}^{+}<|\textbf{k}|<k_{F}^{-}$, 
see Fig. \ref{fig:au-spin-flip}),
the calculated values for the transition probability matrix elements are also 
about twice the ones predicted by the Rashba model.

\section{CONCLUSIONS}
\label{sec:conclusions}

We have analyzed the spin-flip excitations
induced by a time-dependent electric field in the Rashba model textbook example Au(111) surface.
We have considered an ab-initio scheme based on maximally localized Wannier functions, including the full 
spinor structure of the surface states.
Remarkably, the calculated ab-initio spin-flip transition probability
exhibits an appreciable angular and momentum dependence, showing a much more
complex structure than the Rashba model prediction.
An important consequence of this modulation is that the maximum of the calculated spin-flip absorption
rate is about twice the value predicted by the Rashba model.
Thus, even though the Rashba model properly describes
the ground state properties of the surface states, 
these results reveal that it does not fully account for the
dynamical properties.

\section*{ACKNOWLEDGMENTS}
We are grateful to Ivo Souza for 
very helpful discussions.
The authors acknowledge financial 
support from UPV/EHU (Grant No. IT-366-07 and program UFI 11/55), 
the Spanish Mineco (Grant No. FIS2012-36673-C03-01
and FIS2009-12773-C02-01)
and the Basque Government (Grant No. IT472-10),
and thank the Donostia International Physics Center (DIPC)
for providing the computer facilities.

\appendix

\section{SPIN-FLIP TRANSITIONS IN THE RASHBA MODEL}
\label{appendix:rashba}

In this appendix we briefly describe the 
structure of the spin-flip transitions in the Rashba model
for a 2D free-electron-like gas~\cite{rashba},
which is broadly considered as the standard model 
for analyzing the properties of surface states
with spin-orbit interaction~\cite{so-bihlmayer}.
Within this model, electrons are considered as free particles under the action
of a simplified spin-orbit coupling, 
\begin{equation}\label{eq:rashba-so-ham}
\hat{\text{H}}_{\text{R}}=\alpha_{R}\hat{\bm{\sigma}}\cdot(\hat{\textbf{p}}\times \hat{\textbf{z}}).
\end{equation}
Above, $\alpha_{R}$ is the material-dependent Rashba parameter,
$\hat{\textbf{p}}=-i\hat{\boldsymbol{\nabla}}({\textbf{r}})$ is the electron momentum operator
and $\hat{\textbf{z}}$ is a unit vector pointing along the 
surface-perpendicular direction  $z$. 
%
%

The expression of the Hamiltonian in the Rashba model is given by
\begin{equation}\label{eq:rashba-full-ham}
H_{0}=\dfrac{\hat{\textbf{p}}^{2}}{2m^{*}}+\alpha_{R}\hat{\bm{\sigma}}\cdot(\hat{\textbf{p}}\times \hat{\textbf{z}}),
\end{equation}
where $m^{*}$ is the electron effective mass.
For an in-plane electron momentum $\textbf{k}=k(\hat{\textbf{x}}\cos\varphi + \hat{\textbf{y}}\sin\varphi)$, 
the spinor eigenstates of the above Hamiltonian are given by~\cite{so-bihlmayer} 
\begin{equation}\label{eq:rashba-eigenfunc}
\Psi_{\textbf{k},\pm}(\textbf{r})=
\frac{e^{i\textbf{k}\cdot\textbf{r}}}{2\pi}\frac{1}{\sqrt{2}}
\left(\begin{matrix}
  ie^{-i\varphi/2}\\
  \pm e^{i\varphi/2}\\
\end{matrix}\right).
\end{equation}
Above, $\pm$ denote the spin-up and spin-down sub-bands.
The associated momentum dependent 
spin polarization is 
\begin{equation}\label{eq:rashba-spinpol}
\textbf{m}_{\pm}(\textbf{k})=
\Braket{\Psi_{\textbf{k},\pm} |\bm{\sigma}|\Psi_{\textbf{k},\pm}}=
\frac{1}{2}(\hat{\textbf{x}}\sin\varphi \mp\hat{\textbf{y}}\cos\varphi),
\end{equation}
which is perpendicular to both, the electron momentum \textbf{k}
and the surface-perpendicular direction $\hat{\textbf{z}}$.

The energy dispersion corresponding to the 
spinor states of Eq. \ref{eq:rashba-eigenfunc}
is
\begin{equation}\label{eq:rashba-energy}
\epsilon_{k,\pm}=\dfrac{ k^{2}}{2m^{*}} \pm \alpha_{R}k.
\end{equation}
Thus, the eigenstates of the Rashba model Hamiltonian
show an energy spin-splitting which increases linearly with the electron momentum magnitude,
\begin{equation}\label{eq:app-splitting} 
\Delta E_{k}=\epsilon_{k,+}-\epsilon_{k,-}=2\alpha_{R}k.
\end{equation}

Due to the Rashba spin-orbit coupling term of Eq. \ref{eq:rashba-so-ham}, 
the velocity operator,
$\hat{\textbf{v}}= \partial\hat{\text{H}}_{\text{0}}/\partial \hat{\textbf{p}}$
(Eq. \ref{eq:vc}), 
becomes spin-dependent,
\begin{equation}\label{eq:rashba-vel-op}
\begin{split}
& v_{x} = \dfrac{\hat{p}_{x}}{m^{*}} - \alpha_{R}\hat{\sigma_{y}}, \\
& v_{y} = \dfrac{\hat{p}_{y}}{m^{*}} + \alpha_{R}\hat{\sigma_{x}}.
\end{split}
\end{equation}
%
%
Considering the above expressions, the transition matrix elements between
the $\Psi_{\textbf{k},\pm}(\textbf{r})$ states (Eq. \ref{eq:sf-mat-elem-1}),
\begin{equation}\label{eq:rashba-spin-flip-mat-elem-general} 
C^{(\alpha)}_{-+}(\textbf{k})=-\dfrac{e}{2c}
\textbf{A}^{(\alpha)}_{\text{0}}\cdot
\bra{\Psi_{-\textbf{k}}}
 \hat{\textbf{v}}\ket{\Psi_{+\textbf{k}}},
\end{equation}
are directly accessible.  

First, it is worth noting that the spin-diagonal part of the velocity operator
(the canonical contribution $\hat{\textbf{p}}/m^{*}$) does not contribute
to $C^{(\alpha)}_{-+}(\textbf{k})$ due to the orthogonality of the 
$\Psi_{\pm\textbf{k}}(\textbf{r})$ states in spin-basis.
Therefore, the only finite contribution to $C^{(\alpha)}_{-+}(\textbf{k})$ is proportional to the  
Pauli matrices appearing in Eq. \ref{eq:rashba-vel-op}.
We find the following expressions for different light polarizations:
\begin{align}
&C^{(x)}_{-+}(\varphi)=\frac{A_{0}}{2c}
\bra{\Psi_{-\textbf{k}}}
\alpha_{R}\hat{\sigma}_{y}
\ket{\Psi_{+\textbf{k}}}=   i\frac{\alpha_{R}A_{0}}{2c} \sin\varphi, \label{eq:vx-rashba}\\
&C^{(y)}_{-+}(\varphi)=-\frac{A_{0}}{2c}
\bra{\Psi_{-\textbf{k}}}
\alpha_{R}\hat{\sigma}_{x}
\ket{\Psi_{+\textbf{k}}}
=   -i\frac{\alpha_{R}A_{0}}{2c} \cos\varphi, \label{eq:vxyrashba}\\
&C^{(R,L)}_{-+}(\varphi)=\frac{eA_{0}}{2\sqrt{2}c}
\bra{\Psi_{-\textbf{k}}}
\alpha_{R}\hat{\sigma}_{y}\mp
i\alpha_{R}\hat{\sigma}_{x}
\ket{\Psi_{+\textbf{k}}}\nonumber\\
&\;\;\;\;\;\;\;\;\;\;\;\;\;\;\;=   \pm\frac{\alpha_{R}A_{0}}{2\sqrt{2}c} e^{\mp i\varphi} \label{eq:vRLrashba}.
\end{align}
Therefore, the transition matrix elements depend only on the
direction of the electron momentum $\varphi$, but not on the magnitude $k$.
The associated spin-flip transition probability, 
$P^{(\alpha)}_{-+}(\textbf{k})\equiv 
|C_{-+  }^{(\alpha)}(\textbf{k})|^{2}/|\textbf{A}^{(\alpha)}_{0}|^{2}$
(see Eq. \ref{eq:transition-prob}),
is straightforwardly obtained from the above expressions, 
\begin{eqnarray}
\label{eq:rashba-x-mat-elem}
& P_{-+}^{(x)}(\varphi)=
\dfrac{\alpha_{R}^{2}}{4c^{2}}\sin^{2}\varphi, \\
\label{eq:rashba-y-mat-elem}
& P_{-+}^{(y)}(\varphi)=
\dfrac{\alpha_{R}^{2}}{4c^{2}}\cos^{2}\varphi.\\
&P_{-+}^{(R)}=P_{-+}^{(L)}=
\dfrac{\alpha_{R}^{2}}{8c^{2}}.
\label{eq:rashba-RL-mat-elem}
\end{eqnarray}
The last equation shows that the transition probability for
$R$ and $L$ polarized light is identical and independent of 
the electron momentum.

Finally, we have all the elements needed to compute the spin-flip 
absorption rate (Eq. \ref{eq:tr-rate}),
\begin{equation}\label{eq:app-abs-rate1}
\begin{split}
&\Lambda_{-+}^{(\alpha)}(\omega)= 
\frac{\omega}{\mathcal P}
2\pi \int \frac{d^{2}k}{(2\pi)^{2}} 
F_{k}
|C^{(\alpha)}_{-+}(\varphi)|^{2}
\delta(\Delta E_{k}-\omega)\\
&=\dfrac{4}{c|A_{0}|^{2}\omega}
G^{(\alpha)}
\int dk F_{k}
k
\delta(2\alpha_{R}k -\omega),
\end{split}
\end{equation}
where $F_{k}=f_{k,-}-f_{k,+} $
takes into account the Fermi occupation factors,
\begin{equation}
f_{k,\pm}=\left[\exp\left(\dfrac{
k^{2}/2m^{*} \pm \alpha_{R}k
-E_{F}}{k_{B}T}\right)+1\right]^{-1},
\end{equation}
with $E_{F}$ the Fermi energy, $k_{B}$ the Boltzmann constant and $T$ the temperature.
In Eq. \ref{eq:app-abs-rate1}, $G^{(\alpha)}$ denotes the integration of the angular part,
which yields the same result for all light polarizations
(see Eqs. \ref{eq:vx-rashba}$-$\ref{eq:vRLrashba}),
\begin{equation}\label{eq:app-angular-int}
G^{(\alpha)}=\int_{0}^{2\pi}\text{d}\varphi |C^{(\alpha)}_{-+}(\varphi)|^{2}=
\dfrac{|A_{0}|^{2}\pi \alpha_{R}^{2}}{4c^{2}}.
\end{equation} 
Inserting Eq. \ref{eq:app-angular-int} into Eq. \ref{eq:app-abs-rate1}
and integrating over $k$, we finally obtain 
\begin{equation}\label{eq:rashba-absorption-rate}
\Lambda_{-+}^{(\alpha)}(\omega)=
\dfrac{\pi }{4c}\big(f_{\omega,-}-f_{\omega,+}\big).
\end{equation}
The spin-flip absorption rate in the Rashba model, Eq. \ref{eq:rashba-absorption-rate}, turns out to be 
independent of the external field polarization. 
Furthermore, it is almost independent of the Rashba parameter $\alpha_{R}$,
which enters only through the occupation factors. 
Indeed, for $T\rightarrow0$ we have that
\begin{equation}\label{eq:rashba-absorption-rate-T0}
 \lim_{T\rightarrow0}\Lambda_{-+}^{(\alpha)}(\omega)=
 \left\{
\begin{array}{l l}
    \dfrac{\pi }{4c} & \quad \text{\small{if} }
     \omega^{2}/2+\alpha_{R}\omega>E_{F}\\
 & \;\; \text{\small{and }} \omega^{2}/2-\alpha_{R}\omega<E_{F},
    \\
    0 & \quad \text{\small{otherwise}}.
  \end{array} \right.\
\end{equation}
Therefore, we conclude that the parameter $\alpha_{R}$ does not affect the magnitude of the
light absorption rate,
but only the range of frequencies in which light is absorbed.

\section*{References}
\bibliographystyle{apsrev}

\begin{thebibliography}{39}
\expandafter\ifx\csname natexlab\endcsname\relax\def\natexlab#1{#1}\fi
\expandafter\ifx\csname bibnamefont\endcsname\relax
  \def\bibnamefont#1{#1}\fi
\expandafter\ifx\csname bibfnamefont\endcsname\relax
  \def\bibfnamefont#1{#1}\fi
\expandafter\ifx\csname citenamefont\endcsname\relax
  \def\citenamefont#1{#1}\fi
\expandafter\ifx\csname url\endcsname\relax
  \def\url#1{\texttt{#1}}\fi
\expandafter\ifx\csname urlprefix\endcsname\relax\def\urlprefix{URL }\fi
\providecommand{\bibinfo}[2]{#2}
\providecommand{\eprint}[2][]{\url{#2}}

\bibitem[{\citenamefont{Rashba}(1960)}]{rashba}
\bibinfo{author}{\bibfnamefont{E.~I.} \bibnamefont{Rashba}},
  \bibinfo{journal}{Sov. Phys. Solid State} \textbf{\bibinfo{volume}{2}},
  \bibinfo{pages}{1109} (\bibinfo{year}{1960}).

\bibitem[{\citenamefont{Heide et~al.}(2006)\citenamefont{Heide, Bihlmayer,
  Mavropoulos, Bringer, and Bl\"ugel}}]{so-bihlmayer}
\bibinfo{author}{\bibfnamefont{M.}~\bibnamefont{Heide}},
  \bibinfo{author}{\bibfnamefont{G.}~\bibnamefont{Bihlmayer}},
  \bibinfo{author}{\bibfnamefont{P.}~\bibnamefont{Mavropoulos}},
  \bibinfo{author}{\bibfnamefont{A.}~\bibnamefont{Bringer}}, \bibnamefont{and}
  \bibinfo{author}{\bibfnamefont{S.}~\bibnamefont{Bl\"ugel}},
  \bibinfo{journal}{Psi-k Newsletter} \textbf{\bibinfo{volume}{78}},
  \bibinfo{pages}{1109} (\bibinfo{year}{2006}).

\bibitem[{\citenamefont{Pascual
  et~al.}(2004{\natexlab{a}})\citenamefont{Pascual, Bihlmayer, Koroteev, Rust,
  Ceballos, Hansmann, Horn, Chulkov, Bl\"ugel, Echenique
  et~al.}}]{spin-interf-1}
\bibinfo{author}{\bibfnamefont{J.~I.} \bibnamefont{Pascual}},
  \bibinfo{author}{\bibfnamefont{G.}~\bibnamefont{Bihlmayer}},
  \bibinfo{author}{\bibfnamefont{Y.~M.} \bibnamefont{Koroteev}},
  \bibinfo{author}{\bibfnamefont{H.-P.} \bibnamefont{Rust}},
  \bibinfo{author}{\bibfnamefont{G.}~\bibnamefont{Ceballos}},
  \bibinfo{author}{\bibfnamefont{M.}~\bibnamefont{Hansmann}},
  \bibinfo{author}{\bibfnamefont{K.}~\bibnamefont{Horn}},
  \bibinfo{author}{\bibfnamefont{E.~V.} \bibnamefont{Chulkov}},
  \bibinfo{author}{\bibfnamefont{S.}~\bibnamefont{Bl\"ugel}},
  \bibinfo{author}{\bibfnamefont{P.~M.} \bibnamefont{Echenique}},
  \bibnamefont{et~al.}, \bibinfo{journal}{Phys. Rev. Lett.}
  \textbf{\bibinfo{volume}{93}}, \bibinfo{pages}{196802}
  (\bibinfo{year}{2004}{\natexlab{a}}).

\bibitem[{\citenamefont{Strozecka et~al.}(2011)\citenamefont{Strozecka,
  Eiguren, and Pascual}}]{spin-interf-2}
\bibinfo{author}{\bibfnamefont{A.}~\bibnamefont{Strozecka}},
  \bibinfo{author}{\bibfnamefont{A.}~\bibnamefont{Eiguren}}, \bibnamefont{and}
  \bibinfo{author}{\bibfnamefont{J.~I.} \bibnamefont{Pascual}},
  \bibinfo{journal}{Phys. Rev. Lett.} \textbf{\bibinfo{volume}{107}},
  \bibinfo{pages}{186805} (\bibinfo{year}{2011}).

\bibitem[{\citenamefont{LaShell et~al.}(1996)\citenamefont{LaShell, McDougall,
  and Jensen}}]{lashell}
\bibinfo{author}{\bibfnamefont{S.}~\bibnamefont{LaShell}},
  \bibinfo{author}{\bibfnamefont{B.~A.} \bibnamefont{McDougall}},
  \bibnamefont{and} \bibinfo{author}{\bibfnamefont{E.}~\bibnamefont{Jensen}},
  \bibinfo{journal}{Phys. Rev. Lett.} \textbf{\bibinfo{volume}{77}},
  \bibinfo{pages}{3419} (\bibinfo{year}{1996}).

\bibitem[{\citenamefont{Eremeev et~al.}(2012)\citenamefont{Eremeev, Nechaev,
  Koroteev, Echenique, and Chulkov}}]{bi-te-etxenike}
\bibinfo{author}{\bibfnamefont{S.~V.} \bibnamefont{Eremeev}},
  \bibinfo{author}{\bibfnamefont{I.~A.} \bibnamefont{Nechaev}},
  \bibinfo{author}{\bibfnamefont{Y.~M.} \bibnamefont{Koroteev}},
  \bibinfo{author}{\bibfnamefont{P.~M.} \bibnamefont{Echenique}},
  \bibnamefont{and} \bibinfo{author}{\bibfnamefont{E.~V.}
  \bibnamefont{Chulkov}}, \bibinfo{journal}{Phys. Rev. Lett.}
  \textbf{\bibinfo{volume}{108}}, \bibinfo{pages}{246802}
  (\bibinfo{year}{2012}).

\bibitem[{\citenamefont{Ast et~al.}(2007)\citenamefont{Ast, Henk, Ernst,
  Moreschini, Falub, Pacil\'{e}, Bruno, Kern, and Grioni}}]{ast_giant_2007}
\bibinfo{author}{\bibfnamefont{C.~R.} \bibnamefont{Ast}},
  \bibinfo{author}{\bibfnamefont{J.}~\bibnamefont{Henk}},
  \bibinfo{author}{\bibfnamefont{A.}~\bibnamefont{Ernst}},
  \bibinfo{author}{\bibfnamefont{L.}~\bibnamefont{Moreschini}},
  \bibinfo{author}{\bibfnamefont{M.~C.} \bibnamefont{Falub}},
  \bibinfo{author}{\bibfnamefont{D.}~\bibnamefont{Pacil\'{e}}},
  \bibinfo{author}{\bibfnamefont{P.}~\bibnamefont{Bruno}},
  \bibinfo{author}{\bibfnamefont{K.}~\bibnamefont{Kern}}, \bibnamefont{and}
  \bibinfo{author}{\bibfnamefont{M.}~\bibnamefont{Grioni}},
  \bibinfo{journal}{Phys. Rev. Lett.} \textbf{\bibinfo{volume}{98}},
  \bibinfo{pages}{186807} (\bibinfo{year}{2007}).

\bibitem[{\citenamefont{Pascual
  et~al.}(2004{\natexlab{b}})\citenamefont{Pascual, Bihlmayer, Koroteev, Rust,
  Ceballos, Hansmann, Horn, Chulkov, Bl\"ugel, Echenique et~al.}}]{bi100}
\bibinfo{author}{\bibfnamefont{J.~I.} \bibnamefont{Pascual}},
  \bibinfo{author}{\bibfnamefont{G.}~\bibnamefont{Bihlmayer}},
  \bibinfo{author}{\bibfnamefont{Y.~M.} \bibnamefont{Koroteev}},
  \bibinfo{author}{\bibfnamefont{H.-P.} \bibnamefont{Rust}},
  \bibinfo{author}{\bibfnamefont{G.}~\bibnamefont{Ceballos}},
  \bibinfo{author}{\bibfnamefont{M.}~\bibnamefont{Hansmann}},
  \bibinfo{author}{\bibfnamefont{K.}~\bibnamefont{Horn}},
  \bibinfo{author}{\bibfnamefont{E.~V.} \bibnamefont{Chulkov}},
  \bibinfo{author}{\bibfnamefont{S.}~\bibnamefont{Bl\"ugel}},
  \bibinfo{author}{\bibfnamefont{P.~M.} \bibnamefont{Echenique}},
  \bibnamefont{et~al.}, \bibinfo{journal}{Phys. Rev. Lett.}
  \textbf{\bibinfo{volume}{93}}, \bibinfo{pages}{196802}
  (\bibinfo{year}{2004}{\natexlab{b}}).

\bibitem[{\citenamefont{Sakamoto et~al.}(2009)\citenamefont{Sakamoto, Oda,
  Kimura, Miyamoto, Tsujikawa, Imai, Ueno, Namatame, Taniguchi, Eriksson
  et~al.}}]{abrupt}
\bibinfo{author}{\bibfnamefont{K.}~\bibnamefont{Sakamoto}},
  \bibinfo{author}{\bibfnamefont{T.}~\bibnamefont{Oda}},
  \bibinfo{author}{\bibfnamefont{A.}~\bibnamefont{Kimura}},
  \bibinfo{author}{\bibfnamefont{K.}~\bibnamefont{Miyamoto}},
  \bibinfo{author}{\bibfnamefont{M.}~\bibnamefont{Tsujikawa}},
  \bibinfo{author}{\bibfnamefont{A.}~\bibnamefont{Imai}},
  \bibinfo{author}{\bibfnamefont{N.}~\bibnamefont{Ueno}},
  \bibinfo{author}{\bibfnamefont{H.}~\bibnamefont{Namatame}},
  \bibinfo{author}{\bibfnamefont{M.}~\bibnamefont{Taniguchi}},
  \bibinfo{author}{\bibfnamefont{P.~E.~J.} \bibnamefont{Eriksson}},
  \bibnamefont{et~al.}, \bibinfo{journal}{Phys. Rev. Lett.}
  \textbf{\bibinfo{volume}{102}}, \bibinfo{pages}{096805}
  (\bibinfo{year}{2009}).

\bibitem[{\citenamefont{Liu and Chang}(2009)}]{minghao}
\bibinfo{author}{\bibfnamefont{M.-H.} \bibnamefont{Liu}} \bibnamefont{and}
  \bibinfo{author}{\bibfnamefont{C.-R.} \bibnamefont{Chang}},
  \bibinfo{journal}{Phys. Rev. B} \textbf{\bibinfo{volume}{80}},
  \bibinfo{pages}{241304} (\bibinfo{year}{2009}).

\bibitem[{\citenamefont{Iba\~nez Azpiroz et~al.}(2011)\citenamefont{Iba\~nez
  Azpiroz, Eiguren, and Bergara}}]{tlsi111}
\bibinfo{author}{\bibfnamefont{J.}~\bibnamefont{Iba\~nez Azpiroz}},
  \bibinfo{author}{\bibfnamefont{A.}~\bibnamefont{Eiguren}}, \bibnamefont{and}
  \bibinfo{author}{\bibfnamefont{A.}~\bibnamefont{Bergara}},
  \bibinfo{journal}{Phys. Rev. B} \textbf{\bibinfo{volume}{84}},
  \bibinfo{pages}{125435} (\bibinfo{year}{2011}).

\bibitem[{\citenamefont{Ohtsubo et~al.}(2012)\citenamefont{Ohtsubo, Hatta,
  Okuyama, and Aruga}}]{tl-ge111}
\bibinfo{author}{\bibfnamefont{Y.}~\bibnamefont{Ohtsubo}},
  \bibinfo{author}{\bibfnamefont{S.}~\bibnamefont{Hatta}},
  \bibinfo{author}{\bibfnamefont{H.}~\bibnamefont{Okuyama}}, \bibnamefont{and}
  \bibinfo{author}{\bibfnamefont{T.}~\bibnamefont{Aruga}},
  \bibinfo{journal}{Journal of Physics: Condensed Matter}
  \textbf{\bibinfo{volume}{24}}, \bibinfo{pages}{092001}
  (\bibinfo{year}{2012}).

\bibitem[{\citenamefont{Yaji et~al.}(2010)\citenamefont{Yaji, Ohtsubo, Hatta,
  Okuyama, Miyamoto, Okuda, Kimura, Namatame, Taniguchi, and Aruga}}]{scontr}
\bibinfo{author}{\bibfnamefont{K.}~\bibnamefont{Yaji}},
  \bibinfo{author}{\bibfnamefont{Y.}~\bibnamefont{Ohtsubo}},
  \bibinfo{author}{\bibfnamefont{S.}~\bibnamefont{Hatta}},
  \bibinfo{author}{\bibfnamefont{H.}~\bibnamefont{Okuyama}},
  \bibinfo{author}{\bibfnamefont{K.}~\bibnamefont{Miyamoto}},
  \bibinfo{author}{\bibfnamefont{T.}~\bibnamefont{Okuda}},
  \bibinfo{author}{\bibfnamefont{A.}~\bibnamefont{Kimura}},
  \bibinfo{author}{\bibfnamefont{H.}~\bibnamefont{Namatame}},
  \bibinfo{author}{\bibfnamefont{M.}~\bibnamefont{Taniguchi}},
  \bibnamefont{and} \bibinfo{author}{\bibfnamefont{T.}~\bibnamefont{Aruga}},
  \bibinfo{journal}{Nat. Commun.} \textbf{\bibinfo{volume}{1}},
  \bibinfo{pages}{1} (\bibinfo{year}{2010}).

\bibitem[{\citenamefont{Iba\~nez Azpiroz et~al.}(2012)\citenamefont{Iba\~nez
  Azpiroz, Eiguren, Sherman, and Bergara}}]{pbge111-spin-flip}
\bibinfo{author}{\bibfnamefont{J.}~\bibnamefont{Iba\~nez Azpiroz}},
  \bibinfo{author}{\bibfnamefont{A.}~\bibnamefont{Eiguren}},
  \bibinfo{author}{\bibfnamefont{E.~Y.} \bibnamefont{Sherman}},
  \bibnamefont{and} \bibinfo{author}{\bibfnamefont{A.}~\bibnamefont{Bergara}},
  \bibinfo{journal}{Phys. Rev. Lett.} \textbf{\bibinfo{volume}{109}},
  \bibinfo{pages}{156401} (\bibinfo{year}{2012}).

\bibitem[{\citenamefont{Hochstrasser et~al.}(2002)\citenamefont{Hochstrasser,
  Tobin, Rotenberg, and Kevan}}]{hw111-exp}
\bibinfo{author}{\bibfnamefont{M.}~\bibnamefont{Hochstrasser}},
  \bibinfo{author}{\bibfnamefont{J.~G.} \bibnamefont{Tobin}},
  \bibinfo{author}{\bibfnamefont{E.}~\bibnamefont{Rotenberg}},
  \bibnamefont{and} \bibinfo{author}{\bibfnamefont{S.~D.} \bibnamefont{Kevan}},
  \bibinfo{journal}{Phys. Rev. Lett.} \textbf{\bibinfo{volume}{89}},
  \bibinfo{pages}{216802} (\bibinfo{year}{2002}).

\bibitem[{\citenamefont{Eiguren and Ambrosch-Draxl}(2009)}]{hw111-calc}
\bibinfo{author}{\bibfnamefont{A.}~\bibnamefont{Eiguren}} \bibnamefont{and}
  \bibinfo{author}{\bibfnamefont{C.}~\bibnamefont{Ambrosch-Draxl}},
  \bibinfo{journal}{New Journal of Physics} \textbf{\bibinfo{volume}{11}},
  \bibinfo{pages}{013056} (\bibinfo{year}{2009}).

\bibitem[{\citenamefont{Datta and Das}(1990)}]{datta}
\bibinfo{author}{\bibfnamefont{S.}~\bibnamefont{Datta}} \bibnamefont{and}
  \bibinfo{author}{\bibfnamefont{B.}~\bibnamefont{Das}},
  \bibinfo{journal}{Appl. Phys. Lett.} \textbf{\bibinfo{volume}{56}}
  (\bibinfo{year}{1990}).

\bibitem[{\citenamefont{Rashba and Efros}(2003)}]{PhysRevLett.91.126405}
\bibinfo{author}{\bibfnamefont{E.~I.} \bibnamefont{Rashba}} \bibnamefont{and}
  \bibinfo{author}{\bibfnamefont{A.~L.} \bibnamefont{Efros}},
  \bibinfo{journal}{Phys. Rev. Lett.} \textbf{\bibinfo{volume}{91}},
  \bibinfo{pages}{126405} (\bibinfo{year}{2003}).

\bibitem[{\citenamefont{Rashba and Sheka}(1961)}]{rashba-sheka}
\bibinfo{author}{\bibfnamefont{E.~I.} \bibnamefont{Rashba}} \bibnamefont{and}
  \bibinfo{author}{\bibfnamefont{V.}~\bibnamefont{Sheka}},
  \bibinfo{journal}{Sov. Phys. Solid State} \textbf{\bibinfo{volume}{3}},
  \bibinfo{pages}{1357} (\bibinfo{year}{1961}).

\bibitem[{\citenamefont{Golovach et~al.}(2006)\citenamefont{Golovach, Borhani,
  and Loss}}]{PhysRevB.74.165319}
\bibinfo{author}{\bibfnamefont{V.~N.} \bibnamefont{Golovach}},
  \bibinfo{author}{\bibfnamefont{M.}~\bibnamefont{Borhani}}, \bibnamefont{and}
  \bibinfo{author}{\bibfnamefont{D.}~\bibnamefont{Loss}},
  \bibinfo{journal}{Phys. Rev. B} \textbf{\bibinfo{volume}{74}},
  \bibinfo{pages}{165319} (\bibinfo{year}{2006}).

\bibitem[{\citenamefont{Khomitsky et~al.}(2012)\citenamefont{Khomitsky,
  Gulyaev, and Sherman}}]{PhysRevB.85.125312}
\bibinfo{author}{\bibfnamefont{D.~V.} \bibnamefont{Khomitsky}},
  \bibinfo{author}{\bibfnamefont{L.~V.} \bibnamefont{Gulyaev}},
  \bibnamefont{and} \bibinfo{author}{\bibfnamefont{E.~Y.}
  \bibnamefont{Sherman}}, \bibinfo{journal}{Phys. Rev. B}
  \textbf{\bibinfo{volume}{85}}, \bibinfo{pages}{125312}
  (\bibinfo{year}{2012}).

\bibitem[{\citenamefont{Rashba}(2004)}]{PhysRevB.70.201309}
\bibinfo{author}{\bibfnamefont{E.~I.} \bibnamefont{Rashba}},
  \bibinfo{journal}{Phys. Rev. B} \textbf{\bibinfo{volume}{70}},
  \bibinfo{pages}{201309} (\bibinfo{year}{2004}).

\bibitem[{\citenamefont{Henk et~al.}(2004)\citenamefont{Henk, Hoesch,
  Osterwalder, Ernst, and Bruno}}]{au111-exp}
\bibinfo{author}{\bibfnamefont{J.}~\bibnamefont{Henk}},
  \bibinfo{author}{\bibfnamefont{M.}~\bibnamefont{Hoesch}},
  \bibinfo{author}{\bibfnamefont{J.}~\bibnamefont{Osterwalder}},
  \bibinfo{author}{\bibfnamefont{A.}~\bibnamefont{Ernst}}, \bibnamefont{and}
  \bibinfo{author}{\bibfnamefont{P.}~\bibnamefont{Bruno}},
  \bibinfo{journal}{Journal of Physics: Condensed Matter}
  \textbf{\bibinfo{volume}{16}}, \bibinfo{pages}{7581} (\bibinfo{year}{2004}).

\bibitem[{\citenamefont{Liu et~al.}(2008)\citenamefont{Liu, Chen, and
  Chang}}]{PhysRevB.78.195413}
\bibinfo{author}{\bibfnamefont{M.-H.} \bibnamefont{Liu}},
  \bibinfo{author}{\bibfnamefont{S.-H.} \bibnamefont{Chen}}, \bibnamefont{and}
  \bibinfo{author}{\bibfnamefont{C.-R.} \bibnamefont{Chang}},
  \bibinfo{journal}{Phys. Rev. B} \textbf{\bibinfo{volume}{78}},
  \bibinfo{pages}{195413} (\bibinfo{year}{2008}).

\bibitem[{\citenamefont{Henk et~al.}(2003)\citenamefont{Henk, Ernst, and
  Bruno}}]{PhysRevB.68.165416}
\bibinfo{author}{\bibfnamefont{J.}~\bibnamefont{Henk}},
  \bibinfo{author}{\bibfnamefont{A.}~\bibnamefont{Ernst}}, \bibnamefont{and}
  \bibinfo{author}{\bibfnamefont{P.}~\bibnamefont{Bruno}},
  \bibinfo{journal}{Phys. Rev. B} \textbf{\bibinfo{volume}{68}},
  \bibinfo{pages}{165416} (\bibinfo{year}{2003}).

\bibitem[{\citenamefont{Premper et~al.}(2007)\citenamefont{Premper, Trautmann,
  Henk, and Bruno}}]{premper}
\bibinfo{author}{\bibfnamefont{J.}~\bibnamefont{Premper}},
  \bibinfo{author}{\bibfnamefont{M.}~\bibnamefont{Trautmann}},
  \bibinfo{author}{\bibfnamefont{J.}~\bibnamefont{Henk}}, \bibnamefont{and}
  \bibinfo{author}{\bibfnamefont{P.}~\bibnamefont{Bruno}},
  \bibinfo{journal}{Phys. Rev. B} \textbf{\bibinfo{volume}{76}},
  \bibinfo{pages}{073310} (\bibinfo{year}{2007}).

\bibitem[{\citenamefont{Sherman}(2003)}]{sherman_minimum}
\bibinfo{author}{\bibfnamefont{E.~Y.} \bibnamefont{Sherman}},
  \bibinfo{journal}{Physical Review B} \textbf{\bibinfo{volume}{67}},
  \bibinfo{pages}{161303} (\bibinfo{year}{2003}).

\bibitem[{\citenamefont{Blount}(1962)}]{blount}
\bibinfo{author}{\bibfnamefont{E.~I.} \bibnamefont{Blount}},
  \bibinfo{journal}{Solid State Physics} \textbf{\bibinfo{volume}{13}},
  \bibinfo{pages}{305} (\bibinfo{year}{1962}).

\bibitem[{\citenamefont{Wang et~al.}(2006)\citenamefont{Wang, Yates, Souza, and
  Vanderbilt}}]{k-gradients}
\bibinfo{author}{\bibfnamefont{X.}~\bibnamefont{Wang}},
  \bibinfo{author}{\bibfnamefont{J.~R.} \bibnamefont{Yates}},
  \bibinfo{author}{\bibfnamefont{I.}~\bibnamefont{Souza}}, \bibnamefont{and}
  \bibinfo{author}{\bibfnamefont{D.}~\bibnamefont{Vanderbilt}},
  \bibinfo{journal}{Physical Review B} \textbf{\bibinfo{volume}{74}},
  \bibinfo{pages}{195118} (\bibinfo{year}{2006}).

\bibitem[{\citenamefont{Berry}(1984)}]{berry}
\bibinfo{author}{\bibfnamefont{M.~V.} \bibnamefont{Berry}},
  \bibinfo{journal}{Proceedings of the Royal Society of London. A. Mathematical
  and Physical Sciences} \textbf{\bibinfo{volume}{392}}, \bibinfo{pages}{45}
  (\bibinfo{year}{1984}).

\bibitem[{\citenamefont{Lopez et~al.}(2012)\citenamefont{Lopez, Vanderbilt,
  Thonhauser, and Souza}}]{wannier-2012-lopez}
\bibinfo{author}{\bibfnamefont{M.~G.} \bibnamefont{Lopez}},
  \bibinfo{author}{\bibfnamefont{D.}~\bibnamefont{Vanderbilt}},
  \bibinfo{author}{\bibfnamefont{T.}~\bibnamefont{Thonhauser}},
  \bibnamefont{and} \bibinfo{author}{\bibfnamefont{I.}~\bibnamefont{Souza}},
  \bibinfo{journal}{Phys. Rev. B} \textbf{\bibinfo{volume}{85}},
  \bibinfo{pages}{014435} (\bibinfo{year}{2012}).

\bibitem[{\citenamefont{Mostofi et~al.}(2008)\citenamefont{Mostofi, Yates, Lee,
  Souza, Vanderbilt, and Marzari}}]{wannier90}
\bibinfo{author}{\bibfnamefont{A.~A.} \bibnamefont{Mostofi}},
  \bibinfo{author}{\bibfnamefont{J.~R.} \bibnamefont{Yates}},
  \bibinfo{author}{\bibfnamefont{Y.}~\bibnamefont{Lee}},
  \bibinfo{author}{\bibfnamefont{I.}~\bibnamefont{Souza}},
  \bibinfo{author}{\bibfnamefont{D.}~\bibnamefont{Vanderbilt}},
  \bibnamefont{and} \bibinfo{author}{\bibfnamefont{N.}~\bibnamefont{Marzari}},
  \bibinfo{journal}{Computer Physics Communications}
  \textbf{\bibinfo{volume}{178}}, \bibinfo{pages}{685} (\bibinfo{year}{2008}).

\bibitem[{\citenamefont{Wang et~al.}(2007)\citenamefont{Wang, Vanderbilt,
  Yates, and Souza}}]{fermi-surf-wann-wang}
\bibinfo{author}{\bibfnamefont{X.}~\bibnamefont{Wang}},
  \bibinfo{author}{\bibfnamefont{D.}~\bibnamefont{Vanderbilt}},
  \bibinfo{author}{\bibfnamefont{J.~R.} \bibnamefont{Yates}}, \bibnamefont{and}
  \bibinfo{author}{\bibfnamefont{I.}~\bibnamefont{Souza}},
  \bibinfo{journal}{Phys. Rev. B} \textbf{\bibinfo{volume}{76}},
  \bibinfo{pages}{195109} (\bibinfo{year}{2007}).

\bibitem[{\citenamefont{Kokalj}(2003)}]{xcrysden}
\bibinfo{author}{\bibfnamefont{A.}~\bibnamefont{Kokalj}},
  \bibinfo{journal}{Computational Materials Science}
  \textbf{\bibinfo{volume}{28}}, \bibinfo{pages}{155 } (\bibinfo{year}{2003}).

\bibitem[{\citenamefont{Giannozzi et~al.}(2009)\citenamefont{Giannozzi, Baroni,
  Bonini, Calandra, Car, Cavazzoni, Ceresoli, Chiarotti, Cococcioni, Dabo
  et~al.}}]{espresso}
\bibinfo{author}{\bibfnamefont{P.}~\bibnamefont{Giannozzi}},
  \bibinfo{author}{\bibfnamefont{S.}~\bibnamefont{Baroni}},
  \bibinfo{author}{\bibfnamefont{N.}~\bibnamefont{Bonini}},
  \bibinfo{author}{\bibfnamefont{M.}~\bibnamefont{Calandra}},
  \bibinfo{author}{\bibfnamefont{R.}~\bibnamefont{Car}},
  \bibinfo{author}{\bibfnamefont{C.}~\bibnamefont{Cavazzoni}},
  \bibinfo{author}{\bibfnamefont{D.}~\bibnamefont{Ceresoli}},
  \bibinfo{author}{\bibfnamefont{G.~L.} \bibnamefont{Chiarotti}},
  \bibinfo{author}{\bibfnamefont{M.}~\bibnamefont{Cococcioni}},
  \bibinfo{author}{\bibfnamefont{I.}~\bibnamefont{Dabo}}, \bibnamefont{et~al.},
  \bibinfo{journal}{Journal of Physics: Condensed Matter}
  \textbf{\bibinfo{volume}{21}}, \bibinfo{pages}{395502}
  (\bibinfo{year}{2009}).

\bibitem[{\citenamefont{Monkhorst and Pack}(1976)}]{MParck}
\bibinfo{author}{\bibfnamefont{H.~J.} \bibnamefont{Monkhorst}}
  \bibnamefont{and} \bibinfo{author}{\bibfnamefont{J.~D.} \bibnamefont{Pack}},
  \bibinfo{journal}{Phys. Rev. B} \textbf{\bibinfo{volume}{13}},
  \bibinfo{pages}{5188} (\bibinfo{year}{1976}).

\bibitem[{\citenamefont{Corso and Conte}(2005)}]{Dalcorso}
\bibinfo{author}{\bibfnamefont{A.~D.} \bibnamefont{Corso}} \bibnamefont{and}
  \bibinfo{author}{\bibfnamefont{A.~M.} \bibnamefont{Conte}},
  \bibinfo{journal}{Phys. Rev. B} \textbf{\bibinfo{volume}{71}},
  \bibinfo{pages}{115106} (\bibinfo{year}{2005}).

\bibitem[{\citenamefont{Souza and Vanderbilt}(2008)}]{dichroism-souza}
\bibinfo{author}{\bibfnamefont{I.}~\bibnamefont{Souza}} \bibnamefont{and}
  \bibinfo{author}{\bibfnamefont{D.}~\bibnamefont{Vanderbilt}},
  \bibinfo{journal}{Phys. Rev. B} \textbf{\bibinfo{volume}{77}},
  \bibinfo{pages}{054438} (\bibinfo{year}{2008}).

\bibitem[{\citenamefont{Yao et~al.}(2008)\citenamefont{Yao, Xiao, and
  Niu}}]{valleytronics-prb}
\bibinfo{author}{\bibfnamefont{W.}~\bibnamefont{Yao}},
  \bibinfo{author}{\bibfnamefont{D.}~\bibnamefont{Xiao}}, \bibnamefont{and}
  \bibinfo{author}{\bibfnamefont{Q.}~\bibnamefont{Niu}},
  \bibinfo{journal}{Phys. Rev. B} \textbf{\bibinfo{volume}{77}},
  \bibinfo{pages}{235406} (\bibinfo{year}{2008}).

\end{thebibliography}

\end{document}